\documentclass[manuscript,acmtog, 12pt]{acmart}

\renewcommand\footnotetextcopyrightpermission[1]{} 
\usepackage{booktabs} 

\usepackage[ruled]{algorithm2e} 
\usepackage{breqn}

\SetAlFnt{\small}
\SetAlCapFnt{\small}
\SetAlCapNameFnt{\small}
\SetAlCapHSkip{0pt}
\IncMargin{-\parindent}
\usepackage{listings}
\usepackage{color}

\definecolor{dkgreen}{rgb}{0,0.6,0}
\definecolor{gray}{rgb}{0.5,0.5,0.5}
\definecolor{mauve}{rgb}{0.58,0,0.82}

\begin{document}
\title{Taint Tracking for WebAssembly} 
\author{Aron Szanto}
\affiliation{%
  \institution{Harvard University}
  \city{Cambridge}
  \state{MA}
  \postcode{02138}
  \country{USA}}
\author{Timothy Tamm}
\affiliation{%
  \institution{Harvard University}
  \city{Cambridge}
  \state{MA}
  \postcode{02138}
  \country{USA}}
 \author{Artidoro Pagnoni}
\affiliation{%
  \institution{Harvard University}
  \city{Cambridge}
  \state{MA}
  \postcode{02138}
  \country{USA}}
\setcopyright{none}
\settopmatter{printacmref=false}
\begin{abstract}
WebAssembly seeks to provide an alternative to running large and untrusted binaries within web browsers by implementing a portable, performant, and secure bytecode format for native web computation. However, WebAssembly is largely unstudied from a security perspective. In this work, we build the first WebAssembly virtual machine that runs in native JavaScript, and implement a novel taint tracking system that allows a user to run untrusted WebAssembly code while monitoring the flow of sensitive data through the application. We also introduce \emph{indirect taint}, a label that denotes the implicit flow of sensitive information between local variables. Through rigorous testing and validation, we show that our system is correct, secure, and relatively efficient, benefiting from the native performance of WebAssembly while retaining precise security guarantees of more mature software paradigms. 
\end{abstract}
\maketitle
\thispagestyle{empty}
\fontsize{12}{14}\selectfont
\section{Introduction}
As web applications grow in size and complexity, they require users to rely on third-party browser plugins. These large programs offer the capacity to handle heavy computational loads in exchange for bulky and potentially insecure non-native implementations. In the past, growing demand for complex applications like video editing software, 3D games, and scientific programs left both users and developers little choice by way of software models for heavy-duty software on the client side. However, an alternative framework known as WebAssembly (wasm) was recently released as the standard of the future for native high-performance computing in the browser. wasm allows for the compilation of C or C++ code into a novel binary instruction set, which browsers will be configured to execute in a sandboxed virtual environment within their runtime engines. Since wasm code is compiled, optimized, static, has a linear memory model, and does not include built-in automatic garbage collection, it is  20x-40x faster than JavaScript \cite{faq}. And because wasm is intended to run natively, its developers have focused intently on security guarantees that were previously intractable in the face of large, third-party codebases. 

Though wasm is both economical and performant, wide adoption by the community requires, as with all new languages, the development of comprehensive security tools atop it so that code can be checked for safety. One important challenge in security analysis is to monitor the flow of sensitive information through a particular program. In other environments, taint tracking has been deployed as a model for strict bookkeeping of sensitive data \cite{taintdroid}. However, there does not yet exist a platform for taint tracking inside the wasm execution environment. Native wasm taint tracking requires the browser to interpret wasm binary code using client-side (JavaScript) software to track the flow of information at runtime. However, (to our knowledge) there does not exist a JavaScript virtual machine (VM) for wasm. Our contributions are thus twofold: we develop the first JavaScript wasm VM, and atop this framework we institute native binary-code-granularity taint tracking. The remainder of this paper will proceed as follows: Section \ref{background} gives an overview of the WebAssembly technical specification and describes related work in taint tracking. Section \ref{approach} describes our technical approach to building both the JavaScript wasm Virtual Machine and the taint tracking software linked to it. Section \ref{testing} describes the test environment, including three parts: a parallel compilation of C into assembly and WebAssembly with verification of their equivalence via our virtual machine, an extensive suite of taint tracking tasks to validate the correctness of our methods, and a performance evaluation to demonstrate the relative efficiency of our taint tracking implementation. Section \ref{conclusion} concludes and suggests avenues for future work.

\section{Background and Related Work}\label{background}
\subsection{WebAssembly Technical Overview}
WebAssembly is a low-level bytecode format designed to be compiled from C and C++ and run natively in web browsers. In the past, users of complex applications would have to install browser plugins, which are cumbersome and untrusted. WebAssembly allows for native execution of high-performance code within the browser while adhering to strict security guidelines like sandboxed execution and deterministic behavior. Since its development and MVP phase in 2016, WebAssembly has enjoyed quick adoption by major browsers-- an October 2017 estimate put the share of browsers supporting wasm at 61.34\% \cite{caniusewasm}.

WebAssembly's runtime engine is described as a ``structured stack machine'' \cite{wasmsemantics} in that most wasm computations involve a local stack of values, function calls push and pop values from the stack, and control flow is organized into blocks, ifs, and loops. Each binary operation code (opcode) is parsed serially and independently, with the full binary syntax specified as the instantiation of a formal semantics. This allows wasm to define an abstract runtime structure that is hardware-agnostic, allowing for full portability across languages, browsers, operating systems, and machines \cite{wasmWP}. However, in exchange for this flexibility, instructions are of variable length, which complicates interpreter implementations.

WebAssembly bytecode functions are organized into blocks of instructions which are decoded and executed in sequence. An opcode specifies either a control instruction, which may change the state of the program instruction counter (similar to the \%eip value in x86), or a simple instruction, which performs an operation over the values at the top of the stack before pushing the results back onto the stack. wasm opcodes are strongly typed, with each operation specifying an exact datatype (or datatypes) over which it operates. Along with the serial nature of the instruction stream, this guarantees that secure verification of a wasm program can be done in one pass.

WebAssembly's memory model is simple: a linear, contiguous block of memory that is sandboxed away from the stack, local variables, and the runtime engine's memory. This preserves security and ensures simplicity for access (and taint tracking).

Last, wasm does not have direct access to system resources, instead relying on external JavaScript code to pass in data to the virtual environment. However, WebAssembly is able to export data to the runtime environment, meaning that there is a need for a system that monitors wasm code to ensure the proper handling of secure data.
\subsection{Taint Tracking}
Most users regularly use a wide variety of software that (perhaps unbeknownst to the user) has access to sensitive information, including credit card numbers; device hardware data; system, personal, and advertising preferences; and personal identifiers like social security numbers and birth dates. Because software permissions are both coarse-grained and highly opaque, mechanisms for monitoring the flow of sensitive information during the execution of a semi-trusted program are a valuable tool in security analysis.

Taint tracking is a technique that assigns each data object in a program a taint label that contains information about its sensitivity. Taint sources are those that are inherently sensitive (e.g., personal information, IMEI numbers), and their labels are initialized as tainted. As the program executes, taint is transitively propagated between data objects when one object's value is influenced by that of a tainted object. As such, at any point in time the taint tracking engine is able to determine precisely which data are tainted (i.e., either directly or indirectly store sensitive information) and are thus unsafe to transmit to untrusted parties. Many systems for information flow monitoring in previous work are coarse-grained and operate at the emulator level \cite{panorama}, but for our purposes a bytecode-granularity tracking system is required; \cite{taintdroid} describes some of these previous efforts. One recent bytecode-level impelementation is TaintDroid \cite{taintdroid}, a version of taint tracking built atop the Android mobile operating system.  TaintDroid leverages instruction code taint propagation in order to shield users from exposing information to untrusted sources, and to identify applications that act carelessly or maliciously towards users' data privacy.
\\\linebreak We motivate our work by noting that despite the many advantages promised by WebAssembly, its relative youth implies a lack of security tools built atop it. Since JavaScript is the only language that can run natively in a browser, it is essential that there be a JavaScript-based virtual machine that can execute and instrument the wasm bytecode for applications like taint tracking. As such, we build such a machine as a substrate for our taint tracking and other future security software. We follow this with a novel implementation of taint tracking that includes a notion of \emph{indirect taint}, denoting implicit flow of information between variables.
\section{Technical Approach}\label{approach}
\subsection{WebAssembly JavaScript Virtual Machine}
\begin{figure*}[t]
\hrule
\raggedright
\medskip
1. Assert: due to validation, two values of value type t.int32 are on the top of the stack.\\2. Pop the value c2 from the stack.\\3. Pop the value c1 from the stack.\\4. If int32.add(c1,c2) is defined, then:\\\quad \quad Let c be a possible result of computing int32.add(c1,c2).\\\quad \quad Push the value c to the stack.\\5. Else:\\\quad \quad Trap.\\
\medskip
  \hrule
  \caption{Abstract wasm semantics for 32-bit integer addition}
  \label{fig: wasm_32int_add}
\end{figure*}

\begin{figure*}[t]
  \begin{lstlisting}
  if (mod.stack.len() <= 1) {
      return -1;
  }
  c2 = mod.stack.pop();
  c1 = mod.stack.pop();
  if (c1.type != int32_type || c2.type != int32_type) {
      return -1;
  }
  res = (c1.value + c2.value) % Math.pow(2, 32);
  new_var = new Variable(int32_type, res)
  mod.stack.push(new_var);
  \end{lstlisting}
  \caption{Concrete JavaScript syntax for 32-bit integer addition}
  \label{fig: js_32int_add}
\end{figure*}

We organize our VM implementation into two cores: scaffolding and execution. Scaffolding refers to the static building of the virtual environment stipulated by the code's type definitions and other pre-execution rules. Execution is the serial parsing and operation of bytecodes, i.e., running the code.

The largest distributable and executable unit of code in wasm is known as a module. We represent a module as a JavaScript process, meaning that each VM instance contains one module. Within a module, the WebAssembly abstract runtime is organized by \emph{section}, independent components of the engine with global scope and idiosyncratic responsibilities. The sections defined by the specification are: import\footnote{For our proof of concept, we omit an implementation of the import section. Import functionality is analogous to \#include statements in C. Thus, we lose no expresiveness by omitting this, since we could just put all our source code for cross-compilation into one file.}, export, start, global, memory, data, table, elements, function, and code.

The scaffolding routine (\texttt{build_module()} in our implementation) sets up the local environment for the current module, creating JavaScript objects for local memory structures like the runtime stack, instantiating objects for the various sections (e.g., reading and storing static data), and making space for function execution and dynamically-allocated memory. In more detail, scaffolding proceeds as follows: check the magic values and version codes at the beginning of the bytecode to ensure initial validity. Allocate JavaScript objects to hold tables, data, globals, memory, tables, functions, exports, and types (each defined in the formal specification). Starting with the first byte after the header information, read instructions sequentially. For each section code encountered, instantiate JavaScript objects as required by the wasm binary. For example, the scaffolding may encounter a binary code specifying a function section that defines a particular function \texttt{fact} whose type is \texttt{int} $\rightarrow$ \texttt{int}. Our VM would push a \texttt{Type} object with the correct information onto the \texttt{types[]} array for the module. One important responsibility of the scaffolding routine is to load the instruction code into local objects corresponding to each function as defined by the wasm binary. The execution core will then access the instruction code stored in each function object in order to execute the requisite instructions for each function. When the scaffolding routine is finished, the JavaScript process underlying the module will hold local variables that represent the starting state for the execution of the wasm binary.

The execution core is responsible for entering functions, parsing the opcodes within the currently-executing function, and changing the VM's state in accordance with the wasm specification. Formally, the core is tasked with translating the abstract semantics implied by the opcodes it encounters into concrete JavaScript syntax. As a simple example, the abstract semantics define the addition of two 32-bit integers (Figure \ref{fig: wasm_32int_add}) as reading two values from the stack, validating their values, adding them, and then returning the result to the stack. The concrete implementation of these abstract instructions in the JavaScript execution engine is shown in Figure \ref{fig: js_32int_add}. At runtime, the execution core enters an event loop in which it reads and interprets instructions byte by byte, executing the JavaScript code that correspond to the wasm semantics for each instruction. In this way, the execution core is the software bridge between the abstract semantics of wasm and the concrete state changes it implies in the virtual machine.

We close this section by noting that the current implementation does not support floating point operations. In total, the scaffolding and runtime cores, in addition to auxiliary data structures like stacks and variables comprise about 5000 lines of JavaScript code.

\subsection{Taint Tracking}
We approach taint tracking with an eye towards maximal security. As such, we define a taint label per allocable byte in the memory section and for each variable on the stack. We define our taint sources as the set of input parameters to the module, since there is no other path through which information can flow from the outside world to the virtual machine. Specifically, for byte or variable $x$ we represent its taint label as a mapping from an input parameter $y$ to a value in the set $L = \textrm{\{NONE, INDIRECT, DIRECT\}}$, i.e.,  $T(x,y) \in L$. For example, the taint label for variable $x$ in a module with input parameters src1, src2, and src3 might look like:
\begin{lstlisting}
T(x) = {
	 src1: NONE,
    src2: INDIRECT,
    src3: DIRECT
}
\end{lstlisting}
The three cases are defined as follows:\\
\begin{itemize}
  \item $T(x,y) = \textrm{NONE}$ if $x$ is logically isolated from any direct or indirect interaction with the value of $y$, i.e., $x$ has never been tainted by $y$ or $x$ is assigned to a constant or variable untainted by $y$\footnote{This relationship is represented in our implementation by omission in the taint label mapping.}.
  \item $T(x,y) = \textrm{INDIRECT}$ in one of two cases: 
     \begin{itemize}
        \item $x$ is assigned a value in a wasm block (\texttt{block, loop, if}) that includes a conditional expression \texttt{cond ($expr$) \{...\}} where \texttt{cond} is a conditional opcode like \texttt{br_if} or \texttt{br_table}, and any of the variables or bytes in $expr$ are tainted by $y$.
        \item $x$ is assigned a value via $x' = Z[expr]$ where $Z$ is an object on the stack or in memory, $x$ is a variable on the stack and $x=x'$, or else $x$ is a byte affected by the assignment to $x'$ (for example, $x$ is an element in the array $x'$), and any variable in $expr$ is tainted by $y$.
      \end{itemize}
    This means that $x$ is tainted indirectly if $y$ influences the control flow in such a way that $x$ gets a value, or if $y$ influences an array lookup that assigns $x$ a value.
    \item $T(x,y) = \textrm{DIRECT}$ if $x$ was assigned a value $expr$ where $y$ taints any variable in the numeric formula $expr$.
  \end{itemize}

While our taint calculation system is robust, it is not perfect. In particular, there is one rare case that we identify as allowing a variable to escape a would-be tainting operation untainted: if a variable $x$ is assigned a value in a loop body where the loop condition is tainted, and the loop never executes because the condition fails initially, then the variable in the loop body will not be tainted. We allow for this case because attempting to control for it would involve not only tracking the execution of the program as it runs, but also tracking potentially infinite counterfactual executions at each branch to make sure that in no possible execution could $x$ be tainted in the branch body. For example (in C for the sake of simpler reasoning):\\
\begin{lstlisting}
y = (...); // suppose y is tainted and holds value 81
while (y < 42){
	x++;
}
.
.
.
return x;
\end{lstlisting}
Here, if control flow never enters the loop condition, $x$'s value depended indirectly on $y$, though it is never tainted. However, catching this would require cloning the execution state to take both branches; imagining code in which the first branch that was not taken contained many more branches not taken, we might find ourselves in a situation where we have to manage an exponentially large number of copies of the module's state. Symbolic execution systems like KLEE \cite{KLEE} operate in a similar paradigm, cloning its architectural state upon reaching a branch in order to ``take both paths''. However, we call attention to a case in which a potentially tainting operation occurs inside an (infinite) loop; a symbolic execution system might be forced to clone itself infinitely in order to ensure that there is no execution path that would taint any variable. Apart from the theoretical difficulty, this would also be highly computationally expensive, since such a system would have to spin off a separate VM, complete with new local variable and memory environments, for each conditional statement it encounters. We argue that despite this, our taint tracking is superior to many other implementations; for example, TaintDroid does not implement any indirect tainting, let alone the handling of this corner case.

Last, our design does not enforce any alignment or memory access rules in order to ensure that our VM can run arbitrary wasm modules. Because of this, if we were not to keep a taint label for each allocated byte in memory and instead tainted the beginning of each variable, a malicious user would be able to avoid tainting a variable by starting to read it at the memory before its taint label, then discarding the excess bytes at the beginning. Thus, there is no fully secure mechanism for taint tracking in wasm that does not include a complete assignment of taint labels to each active byte in memory. While this design spares no memory expense, it does avoid one common flaw in taint tracking: in many systems (including TaintDroid), arrays are tainted with the union of their constituent elements' taint. Here, since each byte is individually taint tracked, our system ensures a no-false-positive condition.
\section{Testing and Evaluation}\label{testing}
We implement an extensive suite of tests for our system. Each test takes the following form: compile C code for some function into an x86 binary; compile the x86 binary to wasm bytecode; run the wasm code in our system and output the result and the taint; run the x86 binary and compare the result with that from the wasm code; compare the taint output from our system with the (hand-computed) expected taint. In this way, we run a parallel test of a source C function in x86 and on our system to ensure both correctness and taint flow integrity. We implement increasingly complex functions to test our system, and for each we compile them with both no and aggressive compiler optimizations to force our VM to perform each task in a variety of ways. For example, the factorial function compiled with no optimization translates to the wasm engine allocating space in its memory section to compute the result, but the same function compiled with optimizations results in the wasm engine performing the computation using stack variables. We test several programmatic cases, including a variety of branching code (loops, conditionals, etc), recursive functions,  implicit and explicit conversion of datatypes (e.g., casting int64 to int32), and array allocation and random access. Of note are the tests that compute the relatively complex Euler Totient function- we provide two implementations. The first involves a recursive helper function and the second involves a convoluted iterative implementation. We demonstrate that our VM runs them correctly and tracks their taint flawlessly. All of these tests may be run in sequence via the command \texttt{node test.js} from the project root\footnote{It may be necessary to run \texttt{npm install} from the project root to install any missing dependencies}. Note that in two tests we induce the taint tracking error described in Section \ref{approach} to demonstrate its existence only in specific situations.

As part of our test suite, we built an extensive debugging toolset, which allows us to see the state of the stack and memory as each instruction is executed and how taint is being spread. We hope that as our system is used and improved, it will prove helpful to future developers.

We also run several tests to evaluate the performance of the system with and without taint tracking.
\begin{figure}
\includegraphics[width=\columnwidth]{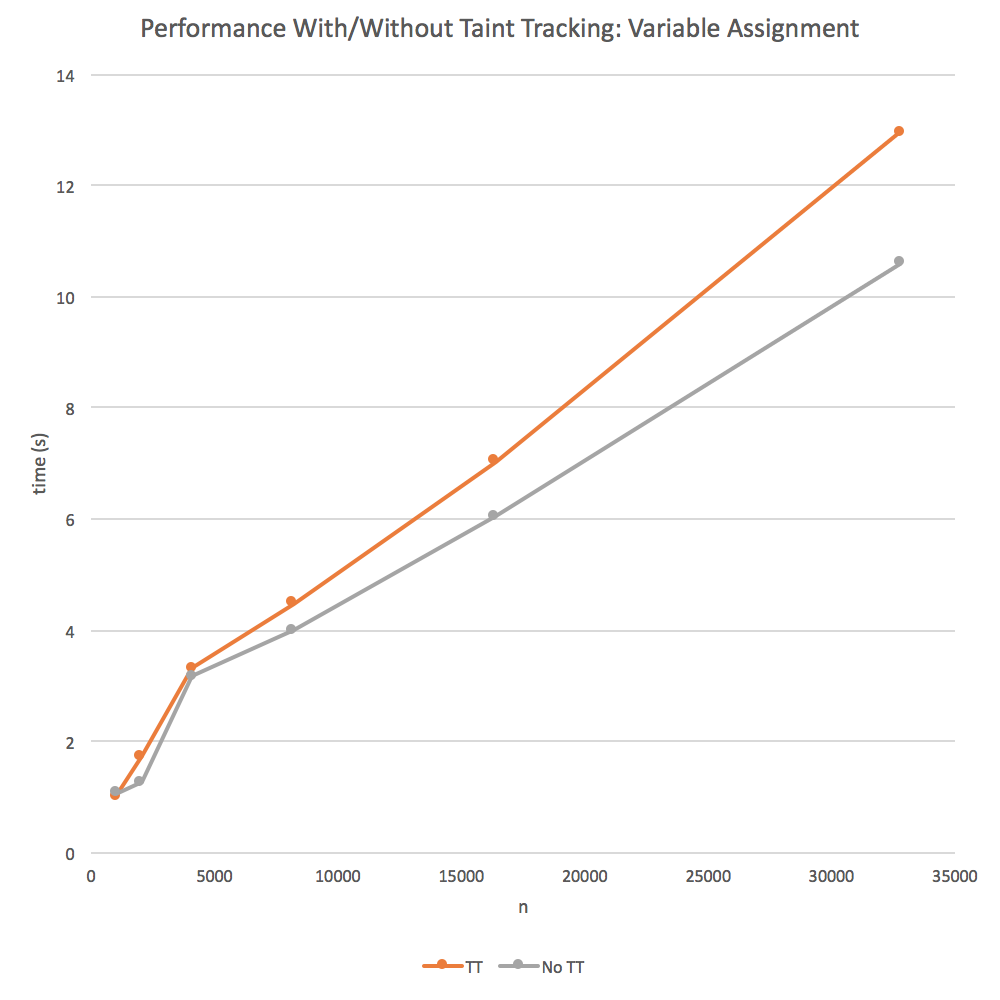}
\caption{Time difference for variable assignment routine with and without taint tracking}
\label{fig: ttperf_va}
\end{figure}
First, we run a simple loop for $n$ iterations, where the input $n$ is tainted. This requires the assignment of a tainted variable $n$ times. We report runtime results for different values of $n$ in Figure \ref{fig: ttperf_va}, finding that taint tracking adds a slight overhead as the number of variable assignments grows into the tens of thousands. This result reflects the simple computational cost of direct taint tracking.
\begin{figure}
\includegraphics[width=\columnwidth]{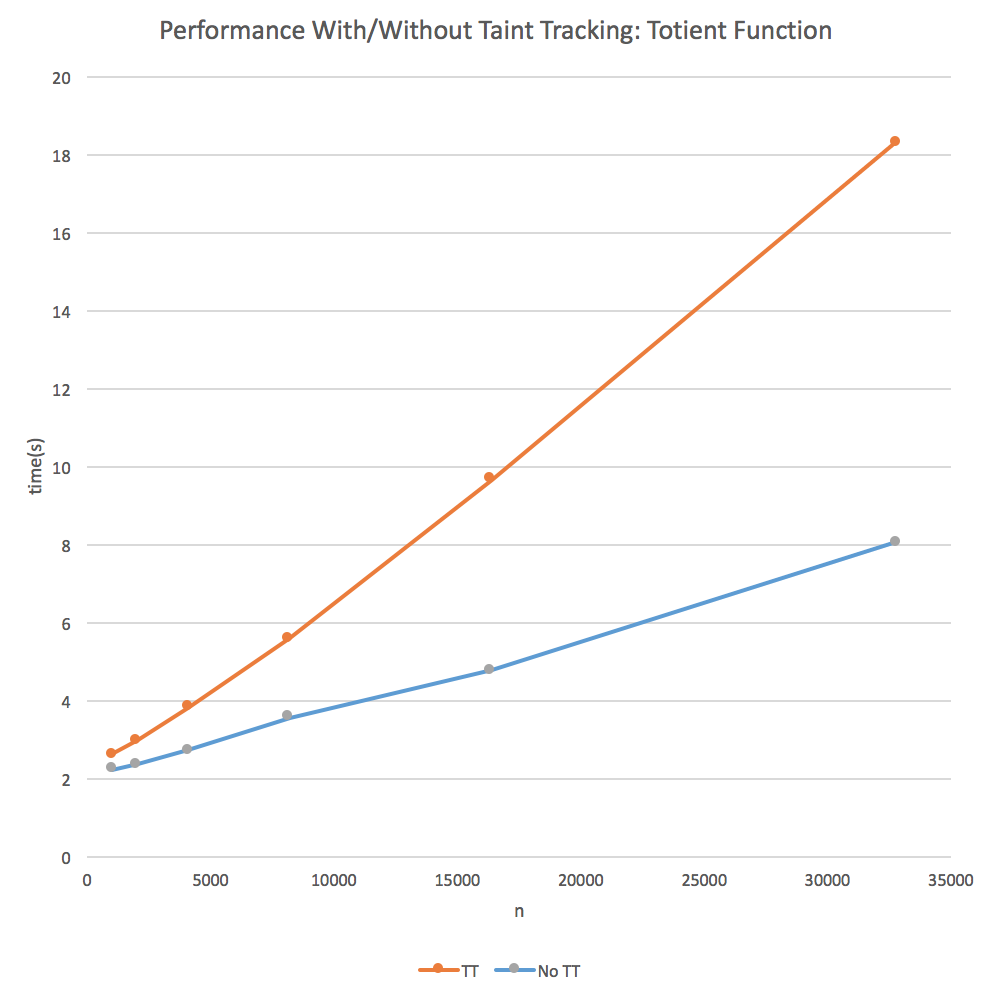}
\caption{Time difference for Totient function routine with and without taint tracking}
\label{fig: ttperf_totient}
\end{figure}

Next, we compute Euler's totient function $\phi$, simulating a working environment in which computing a result is the primary concern, but in which taint tracking is a necessary byproduct. It should be noted that this function induces significant indirect taint transfer. We show results for calculating $\phi(n)$ for several values of $n$ in Figure \ref{fig: ttperf_totient}, finding that the overhead precipitated by the indirect taint transfer represents a more significant performance cost than in the direct taint tracking case. We note, however, that our implementation still guarantees constant-time taint transfer, so no matter the amount of taint propagated, the asymptotic efficiency of any routine will not be affected. Here, though the taint tracked Totient function's runtime curve is steeper than without taint tracking, it still belies a linear-time algorithm.
\begin{figure}
\includegraphics[width=\columnwidth]{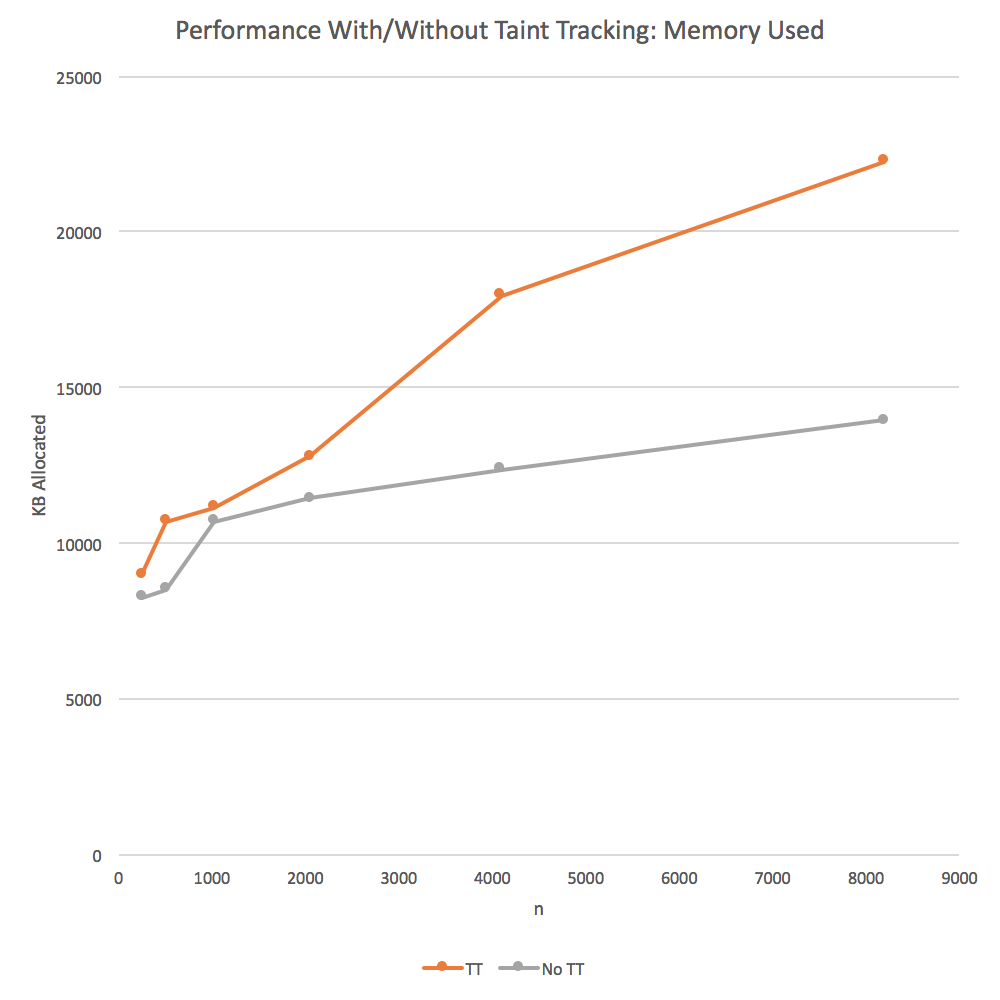}
\caption{Memory overhead from taint tracking}
\label{fig: ttperf_memory}
\end{figure}

Last, we test the memory overhead of taint tracking. Because the objects we allocate are handled by the JavaScript engine, we expect an increase in heap memory usage as we track taint for more variables. We show the results of a test in which we allocate an array of size $n$, then taint each array element. As shown in Figure \ref{fig: ttperf_memory}, memory overhead is, as expected, linear in $n$.

\section{Conclusion}\label{conclusion}
In this paper we make two contributions: a JavaScript virtual machine designed to interpret and run WebAssembly bytecode, and a taint tracking security tool atop the VM to monitor the flow of sensitive information. We describe our design for the scaffolding and execution cores within the wasm engine, and give explicit rules for taint propagation from function inputs to directly and indirectly impacted intermediate and final values. We implement a comprehensive suite of tests that verify the correctness of our JavaScript virtual machine for wasm while validating proper taint tracking procedure. We improve on previous implementations of taint tracking by adding \emph{indirect taint}, a label for a variable whose value is impacted by, though not assigned to, that of a tainted other variable. Finally, we evaluate the performance overhead of our taint tracking implementation, finding that although the tool adds both time and space expense, the excess cost is linearly bounded.

We see several potential avenues of work in the future. First, the VM as written is \emph{almost} complete, omitting some functionality like floating point operations and handling the import section. Future work may complete it for full compliance with the wasm specification. Next, an ambitious extension of our work would be to build a KLEE-like \cite{KLEE} symbolic execution core that not only tracks taint in the current control flow, but taints local variables if they might have been tainted in \emph{any} possible control flow, thus ensuring completeness for indirect taint. Last, the most natural user-facing application of our work is the implementation of a browser add-on that runs taint tracking over untrusted wasm code, demonstrating safe or unsafe handling of a user's private data by a website.

\bibliographystyle{ACM-Reference-Format}
\bibliography{bib}

\end{document}